\documentstyle[proceedings]{JHEP}

\conference{Non-perturbative Quantum Effects 2000}

\newfont{\twelvemsb}{msbm10 scaled\magstep1}
\newfont{\eightmsb}{msbm8}
\newfam\msbfam
\textfont\msbfam=\twelvemsb
\scriptfont\msbfam=\eightmsb
\catcode`\@=11
\def\Bbb{\ifmmode\let\next\Bbb@\else
  \def\next{\errmessage{Use \string\Bbb\space only in math mode}}\fi\next}
\def\Bbb@#1{{\fam\msbfam{{#1}}}}

\newcommand{\be}{\begin{equation}}
\newcommand{\ee}{\end{equation}}
\newcommand{\ba}{\begin{eqnarray}}
\newcommand{\ea}{\end{eqnarray}}
\newcommand{\spz}{\hspace{0.3cm}}
\newcommand{\virg}{\spz,\spz}

\def\d{\delta}

\def\t{\theta}

\newcommand{\la}{\lambda}
\newcommand{\p}{\partial}
\newcommand{\dx}{\partial_x}
\newcommand{\dt}{\partial_t}

\newcommand{\half}{{\textstyle\frac{1}{2}}}

\newcommand{\nn}{\nonumber}
\newcommand{\lt}{\left(}
\newcommand{\rt}{\right)}

\newcommand{\cL}{{\cal L}}

\newcommand{\NP}[1]{Nucl.\ Phys.\ {\bf #1}}
\newcommand{\PL}[1]{Phys.\ Lett.\ {\bf #1}}

\newcommand{\CMP}[1]{Comm.\ Math.\ Phys.\ {\bf #1}}

\newcommand{\PR}[1]{Phys.\ Rev.\ {\bf #1}}
\newcommand{\PRL}[1]{Phys.\ Rev.\ Lett.\ {\bf #1}}
\newcommand{\MPL}[1]{Mod.\ Phys.\ Lett.\ {\bf #1}}

\newcommand{\LMP}[1]{Lett.\ Math.\ Phys.\ {\bf #1}}

\title{Hidden Virasoro Symmetry of the Sine Gordon Theory} 

\author{Davide Fioravanti \thanks{Although this contribution is deeply based
on a series of papers in collaboration with M. Stanishkov, it presents new
results and alternative derivations of results contained in this series of
papers.} \\ S.I.S.S.A.-I.S.A.S. and I.N.F.N. Sez. di Trieste\\ Via Beirut 2-4
34013 Trieste, Italy\\ E-mail: \email{dfiora@sissa.it}}

\abstract{In the framework of the Sine-Gordon (SG) theory we will present
the  construction of a dynamical Virasoro symmetry
which has nothing to do with the space-time Virasoro symmetry of 2D CFT.
Although, it is non-local in the SG field theory,
nevertheless it gives rise to a local action on specific N-soliton solution
variables. These {\it analytic} variables possess a beautiful geometrical
meaning and enter the Form Factor expressions. At the end, we will also give
some preliminary hints about the quantisation.}

\keywords{Integrable System, Symmetries, Conserved Charges}

\begin{document}
\section{Introduction.}
The 2D Sine-Gordon (SG) model, defined by the action:
\be
S={\pi\over\gamma}\int L d^2x \virg  L=(\p_\nu\phi)^2-m^2(cos(2\phi)-1),
\ee
where $\gamma$ is the coupling constant and $m$ is a mass scale,
is one of the simplest massive Integrable Quantum Field
Theories (IQFT's). Nevertheless, it possesses all the fetures peculiar to the
most general IQFT: {\it e.g.} an infinite number of (local) conserved
charges $I_{2n+1}$, $n\in {\Bbb Z}$ in involution, an infrared factorized
scattering theory with solitons, antisolitons and a number of bound
states called ``breathers'' etc.\cite{COL}. Despite this on-shell information, the
off-shell Quantum Field Theory is much less understood. In particular, the
computation of the correlation functions is still a very important
open problem. Actually, some progress toward the direction of an approximative
calculation in the Infra-Red (IR) and Ultra-Violet (UV) regions has been made
recently. For instance, the exact Form-Factors (FF's) of the exponential
fields  $<0|\exp[\alpha\phi(0)]|\beta_1,...,\beta_n>$ were computed \cite{SL}.
This allows one to make predictions about the long-distance behaviour of the
corresponding correlation functions. On the other hand, some efforts have been
made to estimate the short distance behaviour of the theory in the context of
the so-called Conformal Perturbation Theory (CPT) \cite{ALZ}. Good estimates of
interesting physical quantities are possible by combining the previous IR and
UV techniques (\cite{FMS} and reference therein). In addition, the exact
expression for the Vacuum Expectation Values (VEV's) of the exponential fields
(and some descendents) were proposed in \cite{LFZZ}. What still remains
unclear however is the explicit form of the correlation functions, in
particular their intermediate distance behaviour and analytic properties. Some
exact results exist only at the so-called free-fermion point $\gamma={\pi\over
2}$ \cite{MCCOY}.

A radically different approach to the Sine-Gordon theory consists in
searchig for additional infinite-dimentional symmetries and is inspired by the
success the Virasoro symmetry had in the 2D CFT. In fact, it has been shown in
\cite{SBL} that the Sine-Gordon theory possesses an infinite dimensional
symmetry provided by the (level $0$) affine quantum algebra $\widehat sl(2)_q$.
Actually, this symmetry is not useful for determining equations constraining
correlation functions. However, there are some evidences that 
another kind of infinite dimensional symmetry should be present in the
Sine-Gordon theory \cite{LUKY}, a sort of limit of the so-called Deformed
Virasoro  Algebra (DVA) \cite{LUKJAP}.
  
In this talk we present a construction of a new Virasoro symmetry in
the Sine-Gordon theory. Although it will be clear how to implement it in the
general field theory, we are mainly concerned here with the construction of
this symmetry in the case of classical the N-soliton solutions, since -- among
the other resons -- the symmetry in this case is much simpler
realised (in particular it becomes {\it local} contrary to the field theory
case) and the soliton phace space can be quantised in a simpler
manner, as shown in two beautiful papers \cite{BB}, \cite{BBS}.

\section{The Virasoro Symmetry in Field Theory.}
Let us recall the construction of the Virasoro symmetry in the context
of (m)KdV theory \cite{OS, FS}.  It was shown in \cite{FS}, following the
so-called  algebraic approach, that it appears as a generalization of the
ordinary dressing transformations of integrable models. Here we briefly recall 
the main results of this article.
Being integrable, the mKdV system admits a zero-curvature representation:
\be 
[ \dt - A_t , \dx - A_x ] = 0 ,
\label{zcurv}
\ee
where the Lax connections $A_x$ , $A_t$ belong to $A_1^{(1)}$ loop algebra and
contain the field $\phi(x)$ and its derivatives: for instance
\be
A_x= \left(\begin{array}{cc} \phi' & \la \\
                              \la & -\phi' \end{array}\right).
\label{flaxc}
\ee
The usual KdV variable $u(x)$ is connected to the mKdV field $\phi$ by the
Miura transformation:
\be
u={1\over 2}(\phi')^2+{1\over 2}\phi''
\label{miura}
\ee
(we denote by prime the derivative with respect to the {\it space variable} $x$
of KdV). Of great importance in our construction is the solution
$T(x,\lambda)$ of the so called associated linear problem:
\be
(\partial_x-A_x(x,\lambda))T(x,\lambda)=0
\label{aslin}
\ee
which is usually referred to as transfer
matrix. A formal (suitably normalized) solution of (\ref{aslin}) can be easily
found: 
\ba 
&T_{reg}&(x,\la) =  \\
&e^{H\phi(x)}&{\cal P} \exp\lt\la \int_0^xdy
(e^{-2\phi(y)} E +e^{2\phi(y)} F )\rt \nn.
\label{regsol} 
\ea
It is obvious that this solution defines $T(x,\lambda)$ as an infinite series
in positive powers of $\lambda$ with an infinite radius of convergence (we
shall often refer to (\ref{regsol}) as {\it regular expansion}).It is also 
clear from (\ref{regsol}) that $T(x,\la)$ possesses an essential
singularity at infinity where it is governed by the corresponding asymptotic
expansion. It has been derived in detail in \cite{FS0}
\be 
T(x,\la)_{asy}=KG(x,\la)e^{-\int_0^x dy D(y)},
\label{asyexp}
\ee
where $K$ and $G$ and $D$ are written explicitely in \cite{FS0, FS1}. In
particular the matrix 
\be
D(x,\la)=\sum_{i=-1}^{\infty}\la^{-i}d_i(x)H^i \virg  
H= \left(\begin{array}{cc} 1 & 0 \\                                          
                           0 & -1\end{array}\right)  
\ee 
contains the local
conserved densities $d_{2n+2}$ such that $I_{2n+1}=\int_0^L d_{2n+2} dx$.

Obviously, the zero-curvature form (\ref{zcurv}) is preserved by a gauge
transformation for $A_x$
\be
\delta A_x(x,\lambda)=[\t(x,\lambda),{\cal L}]
\label{gauge}
\ee
and a similar one for $A_t$ and we have solely to pay attention to the fact
that the r.h.s. of the previous equation is independent of $\la$ since the
l.h.s is (\ref{flaxc}). Hence, a suitable choice for the gauge parameter
$\theta_n$ goes through the construction of the following object 
\be
Z^X(x,\lambda)=T(x,\lambda)XT(x,\lambda)^{-1}, 
\label{zdress}
\ee
where $X$ is such that
\be
[\dx, X]=0.
\ee
Indeed, it is obvious from the previous definition that it satisfies the 
{\it resolvent condition} 
\be
[\cL,Z^X(x,\la)]=0,
\label{resolv}
\ee
for the first Lax operator $\cL=\partial_x-A_x(x,\lambda)$. Now, this
property means that \be
[\cL,(Z^X(x,\la))_-]=-[\cL,(Z^X(x,\la))_+],
\ee
where the subscript -- (+) means that we restrict the series only to negative
(non-negative) powers of $\la$, and is important for the construction of a
consistent gauge parameter defined as 
\ba
\t^X(x,\la)&=&(Z^X(x,\la))_- \spz or \nn \\  
\t^X(x,\la)&=&(Z^X(x,\la))_+. 
\label{gaugepar}
\ea
Moreover, we have to impose one more consistency condition impied by the
explicit form of $A_x$ (\ref{flaxc}), namely $\delta A_x$ must be diagonal: 
\be 
\d^X A_x=H\d^X \phi' .
\label{selfcons}
\ee
This implies restrictions on the indices of the transformations \cite{FS1}. 
By using $T=T_{reg}$ we obtain the so--called dressing symmetries 
\cite{FS0} and the indices are even for $X=H$ and odd for $X=E,F$. Instead,
by using $T=T_{asy}$ we get for $X=H$ the commuting (m)KdV flows (or mKdV
hierarchy).

At this point we want to make an important observation. Let us consider the KdV
variable $x$ as a {\it space direction} $x_-$ of some more general system (and
$\p_-= \p_x$ as a space derivative). Let us introduce the {\it time} variable
$x_+$ through the corresponding evolution flow 
\be
\p_+ = (\d^E_{-1}+\d^F_{-1}).
\label{timeevol}
\ee
Then, it can be proved \cite{FS0} that the equation of motion
for $\phi$ becomes:
\be
\p_+\p_-\phi=2\sinh(2\phi) 
\ee
or if $\phi\rightarrow i\phi$ 
\be
\p_+\p_-\phi=2sin(2\phi)
\label{sgeq}
\ee
i.e. the Sine-Gordon equation! This observation is very important
for our porpose since it provides an introduction of Sine-Gordon dynamics as a vector field extending the KdV hierarchy flow algebra. Of course, the Hamiltonian given by the dressing charges \cite{FS0} coincides with that previously introduced in \cite{T}. Finally, let us note that
it can be shown that these two kinds of symmetries (regular and asymptotic)
commute with each other. In this sence the non-local regular transformations
provide a true symmetry of the KdV hierarchy. In particullar the $\p_+$
(\ref{timeevol}) evolution of SG equation (as well as the {\it space}
derivative $\p_-$) commutes with the entire hierarchy.

Now, let us explain how the Virasoro symmetry appears in the KdV
system \cite{FS}.
The main idea is that one may dress not only the generators of the underlying 
$A_1^{(1)}$ algebra but also an arbitrary differential operator in the spectral
parameter. We take for example $\lambda^{m+1}\partial_{\lambda}$ which, as it is well 
known, are the vector fields of the diffeomorphisms of the unit circumference
and close a Virasoro algebra. Then we proceed in the same way as above:
\be
Z^V(x,\la)=T(x,\la)\p_\la T(x,\la)^{-1} .
\label{zvir}
\ee
When we consider the asymptotic case, {\it i.e.} we take $T=T_{asy}$ in
(\ref{zvir}), we obtain the non-negative Virasoro flows: {\it e.g.} the first
ones can be written as
\ba
\d_0^V\phi'(x)&=&(x\p+1)\phi'(x), \nn \\
\delta_{2}^V  \phi' &=& 2xa_3'-(\phi')^3+{3\over 4}\phi'''
+2a'_1\int^x_0 d_1, \nn \\ 
\delta_{4}^V  \phi' &=& 2xa_5'+(\phi')^5-{5\over 2}\phi'''(\phi')^2-{27\over
8} (\phi'')^2\phi'+ \nn \\
&+&{5\over 16}\phi^V+2a_3'\int_0^x d_1 +6a_1'\int_0^x d_3,
\label{deltatwo}
\ea
where the $a_j$ are defined in \cite{FS0}. We also constructed the negative
Virasoro flows by taking $T=T_{reg}$ in (\ref{zvir}) \cite{FS}, but here we
are not interested in these flows because only the non-negative part of the
Virasoro flows commute with the light-cone SG dynamics $\p_+$ and $\p_-$.

Now we would like to extend
the construction presented above in (m)KdV theory to the case of Sine-Gordon.
We have already seen in (\ref{timeevol}) how to extend the (m)KdV dynamics to
the SG $\p_+$ flow and for obvious resons we will rename in the following the
KdV variable $u$ with
\be 
u^-=\half(\p_-\phi)^2 +\half \p_-^2\phi.
\ee
Hence, after looking at the symmetric r\^ole that
the derivatives $\p_-$ and $\p_+$ play in the Sine-Gordon equation, we can
obtain the negative (m)KdV flows of the variables $\p_+\phi$ and  
\be
u^+= \half(\p_+\phi)^2 +\half \p_+^2\phi 
\ee
in the same way as
before but with the changes $x_-\rightarrow x_+$ and $\p_-\rightarrow\p_+$
\cite{FS2}. Similarly, we get anhoter half Virasoro algebra by using the same
construction as above but with $x_-$ interchanged with $x_+$. Of course, it is
not obvious at all that the two different halfs will recombine into a unique
Virasoro algebra, but they do! \cite{35} In conclusion, we have found an entire
Virasoro algebra commuting with both light-cone SG dynamics flows $\p_{\pm}$.
Moreover, we prefer to examine in detail the action of this symmetry in the
simple and useful case of soliton solutions.

\section{The Virasoro Symmetry on soliton solution phase space.}
We start with a brief description of the well known soliton solutions of
SG equation (mKdV equation). They are best expressed in terms of the so-called
{\it tau-function}. In the case of N-soliton solution of SG (mKdV) it has the
form: 
\be
\tau(X_1,...,X_N| B_1,...,B_N)=\det(1+V)
\label{tau}
\ee
where $V$ is a matrix:
\be
V_{ij}=2{B_iX_i(x)\over B_i+B_j} \virg i,j=1,...,N.
\label{vmatr}
\ee
The SG (mKdV) field is then expressed as
\be
e^\phi={\tau_-\over \tau_+} ,
\label{phiintau}
\ee
where
\be
\tau_\pm(x)=\tau(\pm X(\dots,t_{-3},t_{-1},t_1,t_3,\dots)|B)
\label{taupm}
\ee
and $X_i(t_{2k+1})$ contains all the {\it times} ({\it e.g.} $t_{-1}=x_+,
t_1=x_-, t_3=t$): 
\be
X_i(x)=x_i\exp(2\sum_{k=-\infty}^{+\infty}B_i^{2k+1} t_{2k+1}).
\label{X}
\ee
The variables $B_i$ and $x_i$ are the parameters describing the solitons:
$\beta_i=\log B_i$ are the so-called rapidities and $x_i$ are related to the
positions. By putting all the negative (positive) {\it times} to zero, we
rediscover the usual mKdV hierarchy (the other with $x_- \rightarrow x_+$).

Our final goal will be the quantization of solitons and of the Virasoro
symmetry. It was argued in \cite{BB} that this is best performed in another
set of variables $\{A_i,B_i\}$. The implicit map from $\{X_i,B_i\}$ to these
new variables is
\be
X_j\prod_{k\ne j}{B_j-B_k\over B_j+B_k}=\prod_{k=1}^N{B_j-A_k\over B_j+A_k} 
\virg j=1,...,N .
\label{analytic}
\ee   
The $\{A_i,B_i\}$ are the soliton limit of certain variables
describing the more general quasi-periodic finite-zone solutions of SG (mKdV)
\cite{Nov}: in that context $B_i$ are the branch points (i.e. define the
complex structure) of the hyperelliptic Riemann surface describing the
solution and $A_i$ are the zeroes of the so-called Baker-Akhiezer function
defined on it. In view of the nice geometrical meaning of these variables they
were called analytical variables in \cite{BB}. In terms of these variables
the tau functions have still a complicated form   
\ba
\tau_+ = 2^N\prod_{j=1}^N B_j\{{\prod_{i<j}(A_i+A_j)\prod_{i<j}(B_i+B_j)\over
\prod_{i,j}(B_i+A_j)}\} \nonumber \\
\tau_- = 2^N\prod_{j=1}^N A_j\{{\prod_{i<j}(A_i+A_j)\prod_{i<j}(B_i+B_j)\over
\prod_{i,j}(B_i+A_j)}\} \nn,
\ea
but, from (\ref{phiintau}), the SG (m-KdV) field enjoys a simple expression    
\be 
e^\phi=\prod_{j=1}^N{A_j\over B_j}.
\label{analiticphi}
\ee
One can verify that, as a consequence, the two components of the
stress-energy tensor (the usual KdV variable) are expressed as 
\ba
u^-&=&\sum_{j=1}^NA_j^2-\sum_{j=1}^NB_j^2 \nn \\ 
u^+&=&\sum_{j=1}^NA_j^{-2}-\sum_{j=1}^NB_j^{-2}.
\label{analitcu}
\ea
In conclusion, we want to restrict the Virasoro symmetry of
SG equation constructed above to the case of soliton solutions. This has been
done in \cite{FS2}, but here we will follow a different path, which
underlines the geometrical meaning of this symmetry. Indeed, our starting
point is the transformation of the rapidities under the Virasoro symmetry.
It can be easily deduced as a soliton limit of the Virasoro action on the
Riemann surface describing the finite-zone solutions \cite{GO}:   
\be 
\d_{2n}B_i=B_i^{2n+1} \virg n\ge 0.
\label{btransf}
\ee
Similarly, for negative transformations we obtain \cite{FS2} 
\be 
\d_{-2n}B_i=-B_i^{-2n+1} \virg n > 0
\label{negb}
\ee
where we have put an additional -- sign in the r.h.s. for preserving the
self-consistency of the construction. In other words, the Virasoro action
changes the complex structure. What remains is to obtain the transformations of
the $A_i$ variables. In \cite{FS2} we have deduced them from the
transformations of the fields $\d_{2n}\phi$, $\d_{2n}\phi'$, $\d_{2n}u$ {\it
etc.} (\ref{deltatwo}) restricted to the soliton solutions using
(\ref{analiticphi},\ref{analitcu}). Instead, in this talk we will show that
those are consequences of (\ref{btransf}) and (\ref{negb}) if applied to the
implicit map (\ref{analytic}) by using the explicit expression of $X_j$ in
terms of $B_i$ (\ref{X}). The problem is simplified by the fact that
the Virasoro algebra is freely generated, {\it i.e.} we need to compute only
the $\d_0$, $\d_{\pm 2}$ and $\d_{\pm 4}$ transformations, since the higher
vector fields are then obtained by commutation. As shown in \cite{FS2},
although the transformations of $\p_{\pm}\phi$ and $u^{\pm}$ in the SG theory
are quasi-local, since they contain certain indefinite integrals, the
corresponding integrands become total derivatives when restricted to the
soliton solutions. Therefore  the Virasoro transformations become {\it local}
when restricted to the soliton solutions! We feel more natural and more
compact to express the Virasoro action on $A_i$ by using the {\it equation of
motions} of $A_i$ derived from (\ref{X}) and (\ref{analytic}), for instance:
\ba
\d_{-1}A_i&=&\p_+A_i=\prod_{j=1}^N{A_i^2-B_j^2)\over B_j^2}\prod_{j\ne
i}{A_j^2\over (A_i^2-A_j^2)}, \nn \\
\d_1 A_i&=&\p_-A_i=\prod_{j=1}^N(A_i^2-B_j^2)\prod_{j\ne
i}{1\over(A_i^2-A_j^2)}, \nn \\
{1\over 3}\d_3A_i &=& (\sum_{j=1}^N B_j^2-\sum_{k\ne i}A_k^2)\p_-A_i , \nn \\ 
{1\over 5}\d_5A_i &=& (\sum_{j=1}^N B_j^4-\sum_{k\ne i}A_k^4)\p_-A_i - \nn \\
&-&\sum_{j\ne i}(A_i^2-A_j^2)\p_-A_i\p_-A_j.
\label{kdvonsol}
\ea
In conclusion, the calculation is
straightforward but quite tedious and we present here only the final result:
\ba
&\d_{-2}&A_i = \frac{1}{3} x_+\d_{-3}A_i
-A_i^{-1}-(\sum_{j=1}^NA_j^{-1})\p_+A_i- \nn \\
&-&x_-\p_+A_i \nn \\ 
&\d_{-4}&A_i = \frac{1}{5} x_+\d_{-5}A_i
-A_i^{-3}-  \\            
&-&\{\sum_{j\neq i}^N{1\over A_i}({1\over
 A_i^2}-{1\over A_j^2})+\sum_{j=1}^N{1\over A_j}\sum_{k=1}^N{1\over
 B_k^2}\}\p_+A_i- \nn \\
&-&x_-\d_{-3}A_i
\label{result1}
\ea
and 
\ba 
\d_0 A_i &=& (x_-\p_- -x_+\p_+ +1)A_i , \nn \\
\d_2 A_i &=&{1\over 3}x_-\d_3A_i+A_i^3-(\sum_{j=1}^N A_j)\p_-A_i-x_+\p_-A_i,
\nonumber \\ 
\d_4 A_i &=&{1\over 5}x_-\d_5A_i+A_i^5-\{\sum_{j\ne
i}A_i(A_i^2-A_j^2)+ \nn \\
&+&\sum_{j=1}^NA_j \sum_{k=1}^N B_k^2\}\p_-A_i-x_+\d_3A_i.
\label{result2}
\ea

At this point, two important checks are are relevant. The firts concerns 
the commutation of the $\d_{2m}$ (with $m\in\Bbb Z$) acting on $A_i$ with the
light-come SG flow $\p_{\pm}$. This check admits a positive answer. The second 
is verifying the
algebra of the $\d_{2m}$ (with $m\in\Bbb Z$) acting on $A_i$ and is non trivial
because we have derived all the transformations (\ref{result1}) and
(\ref{result2}) from the Virasoro algebra on $B_i$ written in (\ref{btransf})
and (\ref{negb}). The action on $A_i$ is again a representation of the
centerless Virasoro algebra:  
\be 
[\d_{2n},\d_{2m}]A_i=(2n-2m)\d_{2n+2m} A_i \virg n,m\in {\Bbb Z}.
\label{primevir} 
\ee

\section{Comments about quantisation.}
Of course, we are interested in the quantum Sine-Gordon
theory. In the case of solitons there is a standard procedure, the 
canonical quantisation of the N-soliton solutions. In fact, let us introduce,
following \cite{BB}, the canonically conjugated variables to the {\it
analytical variables} $A_i$:
\be
P_j=\prod_{k=1}^N{B_k-A_j\over B_k +A_j} \virg j=1...,N.
\ee
In these variables one can perform the canonical quantisation
of the N-soliton system introducing the deformed commutation relations between
the operators $A_i$ and $P_i$ :
\ba
P_jA_j &=& q A_jP_j , \nn \\
P_kA_j &=& A_jP_k \spz for \spz k\ne j ,
\ea
where $q=\exp (i\xi)$, $\xi={\pi\gamma\over \pi-\gamma}$. It is very
intriguing to understand how the Virasoro symmetry is deformed after the
quantisation!

\acknowledgments{The author thanks with pleasure his collaborator about this
research field, Marian Stanishkov. Moreover he thanks 
the I.N.F.N.--S.I.S.S.A. for financial support. This work has been realised
through partial financial support of TMR Contract ERBFMRXCT960012.}

\end{document}